\documentclass{iaus}
\usepackage{graphicx}
\usepackage{AMSsymb}

\title[Faint sub-L* galaxies at $z$$\sim$2] 
{The build-up of mass in\\ UV-selected sub-L* galaxies at $z$$\sim$2}

\author[Marcin Sawicki]   
{Marcin Sawicki}

\affiliation{Saint Mary's University, Halifax, Canada \\
sawicki@ap.smu.ca}

\pubyear{2011}
\volume{277}  
\pagerange{1--4}
\jname{Tracing the Ancestry of Galaxies on the Land of Our Ancestors}
\editors{C.\ Carignan, K.C.\ Freeman, \& F.\ Combes, eds.}
\begin{document}

\maketitle

\begin{abstract}
Broadband spectral energy distribution (SED) fitting is used to study a deep sample of UV-selected sub-L* galaxies at $z$$\sim$2. They are found to be less dusty than L* galaxies, and to contribute more mass to the cosmic mass budget at this epoch than is inferred from shallower high-$z$ surveys. Additionally, SFRs are found to be proportional to stellar masses over three orders of magnitude in mass; this phenomenon can be explained by assuming that new stars form out of gas that co-accretes along with dark matter onto the galaxies' dark matter halos, a scenario that naturally leads to SFRs that gradually increase with time. 

\end{abstract}

\firstsection 
\section{Introduction}

 For over a decade now it has been possible to directly observe normal galaxies at epochs when the Universe was only a fraction of its present age. However, our current understanding of high-redshift galaxies is based largely on luminous objects. This focus has in part been motivated by theoretical interests in massive systems as a means to test hierarchical galaxy formation models, but much of the problem is also practical: because high-$z$ galaxies are faint and thus expensive to observe, most studies to date have focused on relatively luminous objects around the L* ÒkneeÓ in the luminosity function. While economical, this approach has limited us to what are, in fact, quite rare objects. The $\sim$L* ($\sim$M*) galaxies that have been the focus of most studies so far are rare indeed, as can be appreciated in, e.g., the $z\sim2$ luminosity and stellar mass functions (LF, SMF; Fig. 2b). We thus lack good understanding of the bulk of the high-$z$ galaxy population. This work (the full details of which are given in \cite[Sawicki 2011a]{Sawicki 2011a}) aims to redress part of this deficiency by investigating the stellar populations of sub-L* galaxies at $z$$\sim$2.

\section{Data and SED fits}

This study uses HST optical ($U_{300}B_{450}V_{606}I_{814}$) and IR ($J_{110}H_{160}$) images of the Hubble Deep Field (HDF), with object finding and photometry done using SExtractor (\cite[Bertin \& Arnouts 1996]{Bertin-Arnouts-1996}).  The data reach $>$90\% completeness at $V_{606}$=28.8.  High-$z$ galaxies are chosen using the \cite[Steidel et al.\ (1996)]{Steidel-etal-1996} HDF-specific color-color criteria which are good at selecting galaxies at $z$$\gtrsim$1.8. A further 1.8$\leq$$z_{phot}$$\leq$2.6 photometric redshift cut is applied both to guard against potential low-$z$ interlopers and to eliminate galaxies at $z$$>$2.6, where stellar masses cannot be estimated well because the 4000\AA\ break redshifts beyond the available wavelength coverage. The sample then consists of 91 objects with $<$$z$$>$=2.3, and reaches $\sim$4~mag below L*. 

Spectral energy distribution (SED) fitting is performed using the SEDfit package (\cite[Sawicki \& Yee 1998]{Sawicki-Yee-1998}; \cite[Sawicki 2011b]{Sawicki-2011b}).  For consistency with the brighter, L$\sim$L* $z$$\sim$2.3 BX sample of \cite[Shapley et al.\ (2005)]{Shapley-etal-2005}, constant SFR models with the Salpeter IMF, taken from the \cite[Bruzual \& Charlot (2003)]{Bruzual-Charlot-2003} library are used, and these are attenuated with \cite[Calzetti et al.\ (2000)]{Calzetti-etal-2000} starburst-type dust.

\section{Results}

\subsection{Dust}

Figure 1 shows the dependence of color excess, $E(B-V)$, on UV luminosity and on galaxy stellar mass. Fig.~1(a) shows a trend between luminosity and reddening: sub-L* galaxies appear to have less dust than do their $\sim$L* counterparts. More than half the UV photons produced by a typical M*+3 (M*$\sim$$10^9$$M_{\odot}$) BX galaxy escape into intergalactic space, in contrast to $\sim$L* galaxies from which only about 1 in 5 UV photons escape absorption by dust. Sub-L* BX galaxies are thus relatively naked in the UV and as such they are important contributors to the UV luminosity density budget at $z$$\sim$2. 

The data also show that more massive $z$$\sim$2 galaxies tend to be more dusty than lower-mass objects (Fig.~1b).  In contrast to UV luminosity, which can vary rapidly in response to a fluctuating SFR, stellar mass is likely a more stable quantity.  Consequently, it seems plausible that the underlying reddening trend is with stellar mass rather than UV light.

\begin{figure}[t]
\begin{center}$
\begin{array}{ccc}
 \includegraphics[width=2.0in]{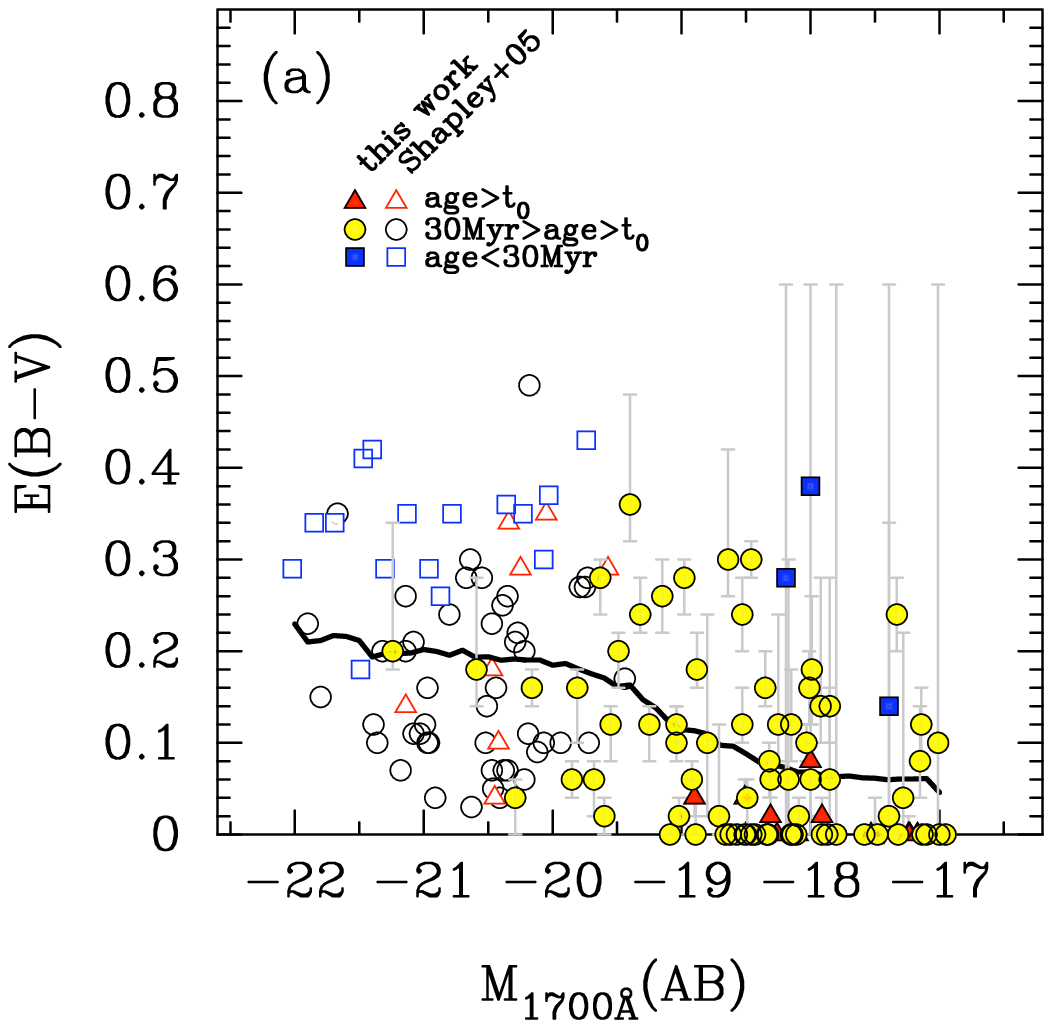} 
 \hspace{1cm}
  \includegraphics[width=2.06in]{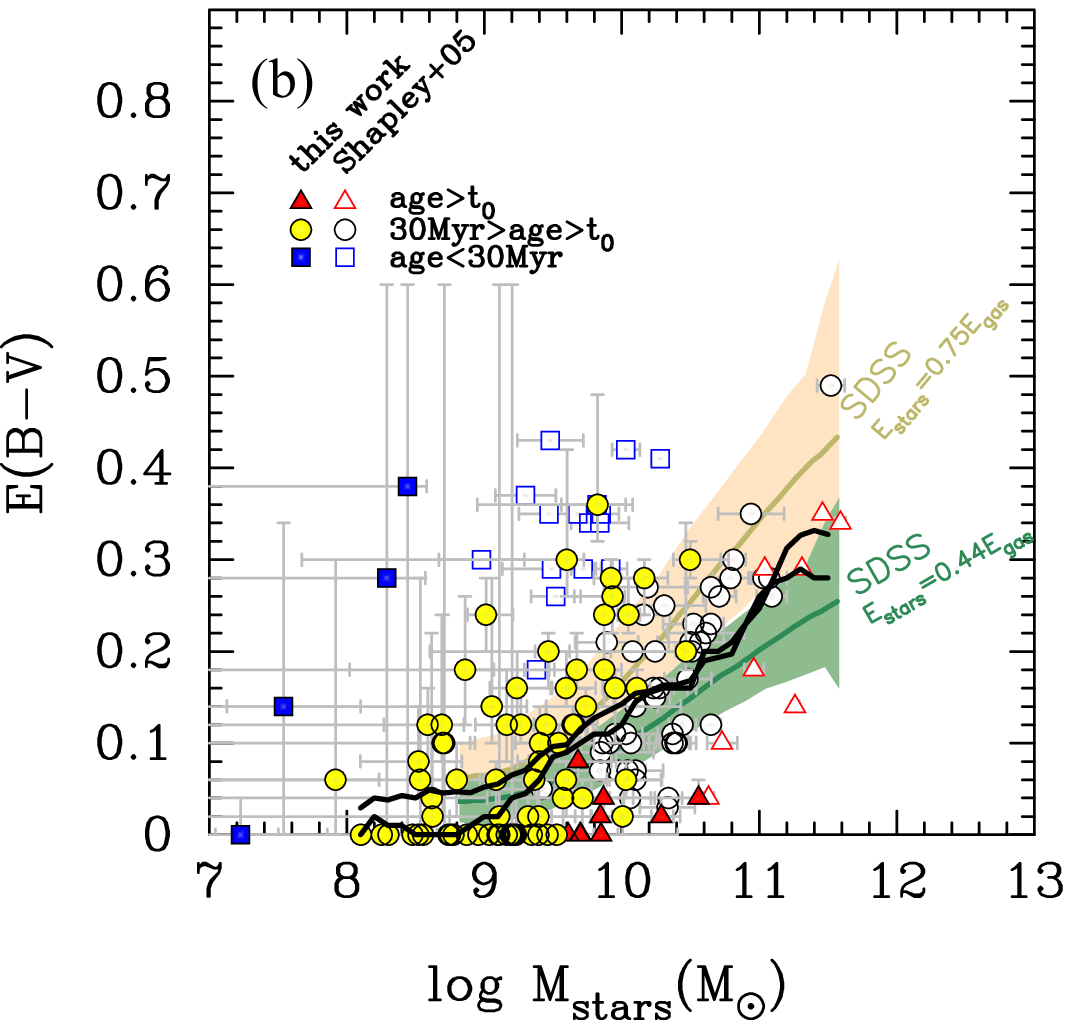} 
\end{array}$
 \caption{Dust properties as a function of UV luminosity {\it (left)} and stellar mass {\it (right)}. The solid black curves show the boxcar-smoothed mean and median values. In the right panel, the colored solid lines and shaded regions show two implementations of the local reddening-stellar mass relation found for SDSS galaxies by \cite[Garn \& Best (2010)]{Garn-Best-2010}. }
\label{fig-dust}
\end{center}
\end{figure}

\subsection{Stellar mass function and stellar mass density of the Universe}

Figure 2(a) shows a correlation between stellar mass and UV luminosity density among $z$$\sim$2 BX galaxies. The observed correlation is fit as $\log (M_{stars}/M_\odot) = 0.68 - 0.46 M_{1700}$, which can be used to convert a UV LF into a stellar mass function.  This conversion is done in Fig.~2(b), which shows several different determinations of the $z$$\sim$2 UV LF, with the top+right axes recasting them as stellar mass functions. From Fig.~2(b) it is clear that the low-mass end of the stellar mass function of UV-selected galaxies at $z$$\sim$2 is steep:  there are many more low-mass galaxies than high-mass ones at $z$$\sim$2.

Figure~3(a) shows that the low-mass end of the SMF at $z$$\sim$2 is steeper than predicted by extrapolations of shallower surveys:  there are more low-mass galaxies than such extrapolations would predict. These numerous low-mass galaxies contribute significantly to the stellar mass density at $z$$\sim$2.3, increasing its value by a factor of $\sim$1.5 compared to extrapolations from higher-mass data alone (Fig.~3b).  This upward revision implies that by $z$$\sim$2.3 the Universe had already made $\sim$20--30\% of its present-day stellar mass, as compared to only $\sim$15\% that the extrapolations of the shallower surveys would suggest. Thus, apparently the build-up of stellar mass in the universe has proceeded somewhat more quickly than previously thought.

\begin{figure}[t]
\begin{center}$
\begin{array}{ccc}
 \includegraphics[width=1.775in]{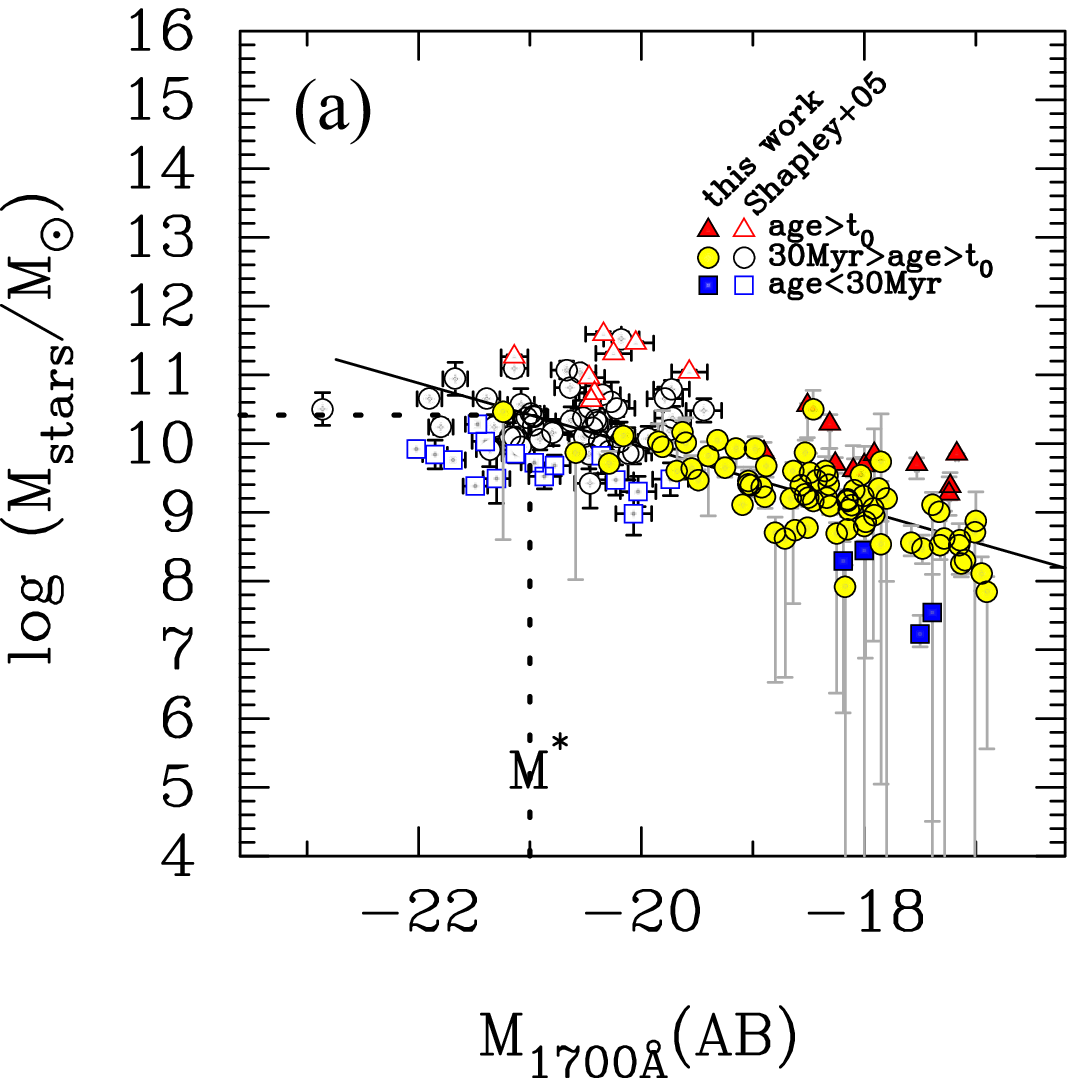} 
 \hspace{1cm}
 \includegraphics[width=3in]{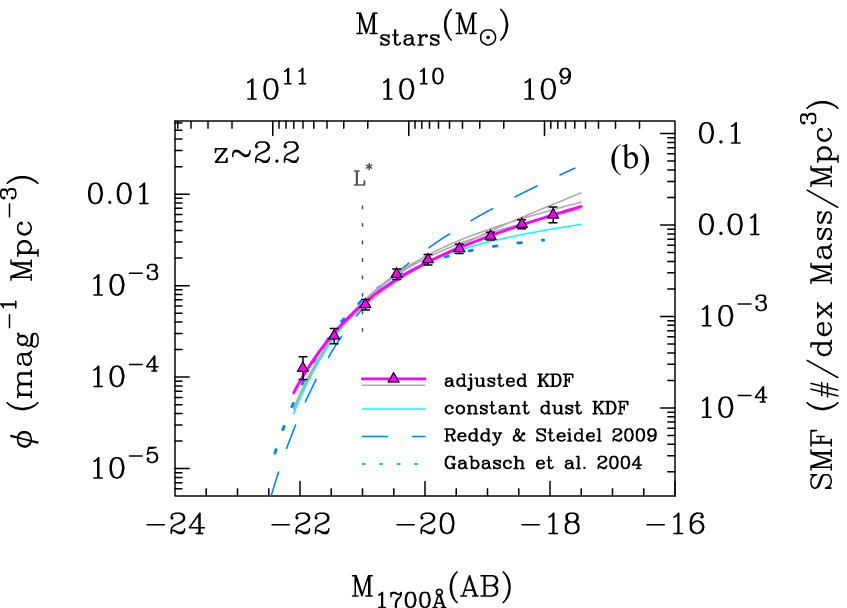} 
\end{array}$
 \caption{ The linear relation between stellar mass and UV luminosity (left panel) can be applied to the UV LF (right panel) to obtain the stellar mass function. In the right panel several UV LFs are shown, including the original KDF one (\cite[Sawicki \& Thompson 2006]{Sawicki-Thompson-2006}) and the KDF LF after the $V_{eff}$ has been adjusted for luminosity-dependent dust (\cite[Sawicki 2011a]{Sawicki-2011a}).}
   \label{fig-dust}
\end{center}
\end{figure}

\begin{figure}[t]
\begin{center}$
\begin{array}{ccc}
 \includegraphics[width=2.18in]{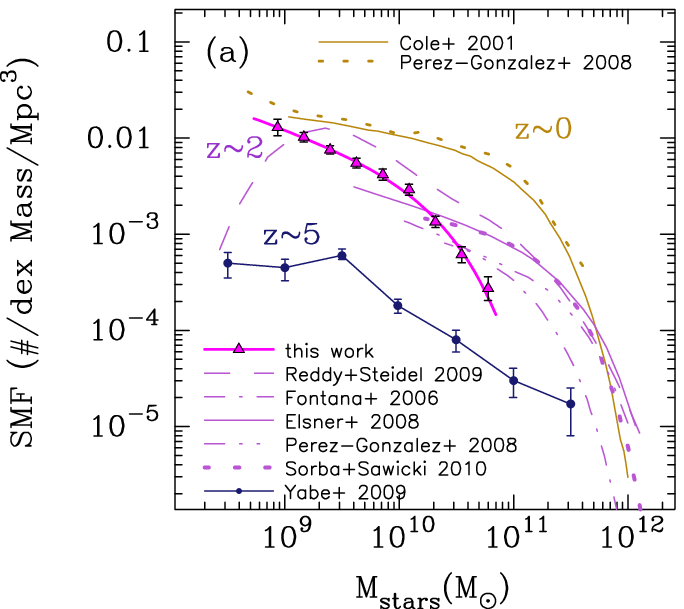} 
\hspace{1cm}
 \includegraphics[width=2in]{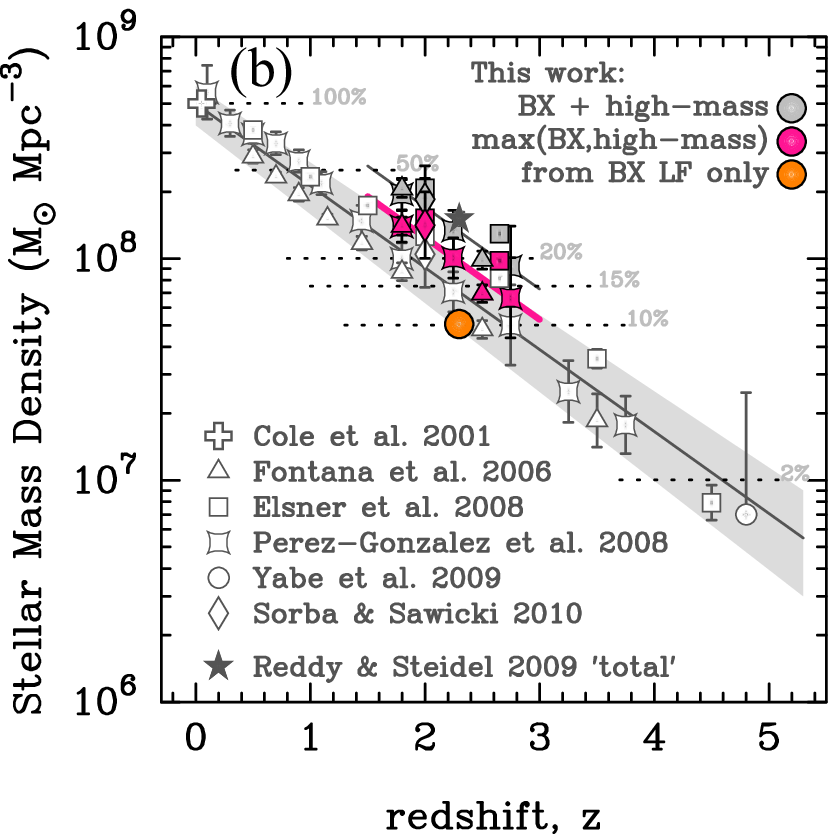} 
\end{array}$
 \caption{{\it Left:} Comparison of stellar mass functions from the literature.  Extrapolations of the shallower SMFs at $z\sim 2$ predict smaller numbers of low-mass galaxies than are found by the two deeper SMFs (present work and that by \cite[Reddy \& Steidel 2009]{Reddy-Steidel-2009}).  Note that the SMF of \cite[Reddy \& Steidel (2009)]{Reddy-Steidel-2009} uses the SFR-$L_{UV}$ conversion of \cite[Sawicki et al.\ (2007)]{Sawicki-etal-2007}, which is essentially identical to that used in the present work. {\it Right:} The stellar mass density of the universe.  The lower filled circle shows the contribution from integrating the stellar mass function in Fig.\ 2(b); the upper filled symbols show the various $z$=1.8--3 results adjusted for the contribution of low-mass BX galaxies.  The main result is that, once the numerous low-mass galaxies are taken into account, the stellar mass density appears to be $\sim$1.4--2 times higher at $z \sim 2$ than would have been thought on the basis of the shallower surveys. 
 }
   \label{fig-dust}
\end{center}
\end{figure}

\subsection{The SFR-M$_{stars}$ relation and mass growth}

\begin{figure}[t]
\begin{center}
 \includegraphics[width=2.3in]{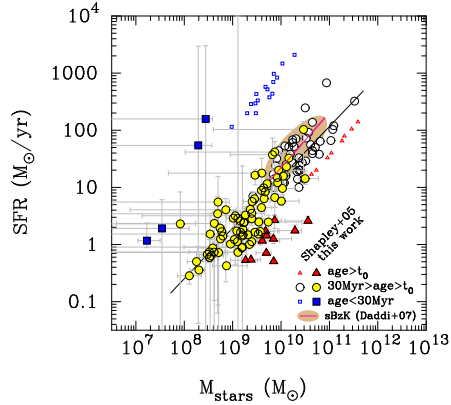} 
 \caption{Star formation rate as a function of stellar mass. 
 }
   \label{fig-dust}
\end{center}
\end{figure}

As Fig.\ 4 shows, there exists, over three orders of magnitude in mass, a relation between SFR and stellar mass, with 
$\log (SFR / M_\odot yr^{-1}) = 0.89 \log ({M_{stars}/M_\odot}) - 7.69.$ The existence of this relation poses the challenge:  how do BX galaxies ``know" that they should have SFR $\propto$ $M_\odot$, or, in other words, SFR $\propto$ $\int$SFR$(t)dt$ ? 

A simple model can account for this phenomenon (details in Sawicki 2011a):  Start by assuming that SFRs are proportional to baryon accretion rates, which in turn are proportional to dark matter accretion rates.  Since in the $\Lambda$CDM paradigm DM accretion rates scale with halo mass, it then follows that both SFRs and stellar masses will correlate via the masses of the underlying DM halos.  Based on the scaling of halo accretion rates with mass, the model successfully accounts for the slope of the observed BX galaxy  SFR-$M_{stars}$ correlation, and finds that the instantaneous star formation efficiency (fraction of accreting baryons that condenses into stars) is low ($\sim$1\%). The key ingredient of this model --- that SFRs are governed by the (continually growing) DM masses of galaxies --- has the potential to naturally account for the increasing star formation histories that have recently been suggested for high-$z$ galaxies (\cite[Lee et al.\ 2010]{Lee-etal-2010}; \cite[Maraston et al.\ 2010]{Maraston-etal-2010}).





\begin{thebibliography}{}

\bibitem[Bertin \& Arnouts (1996)]{Bertin-Arnounts-1996}
{Bertin, E. \& Arnouts, S.} 1996, \textit{A\&A}, 117, 393

\bibitem[Bruzual \& Charlot (2003)]{Bruzual-Charlot-2003}
{Bruzual, G. \& Charlot, S.} 2003, \textit{MNRAS}, 344, 1000

\bibitem[Calzetti et al. (2000)]{Calzetti-etal-2000} 
{Calzetti, D., et al.}  2000, \textit{ApJ}, 533, 682 

\bibitem[Cole et al.\ (2001)]{Cole-etal-2001} 
{Cole, S., et al.} 2001, \textit{MNRAS},  326, 255

\bibitem[Daddi et al.\ (2007)]{Daddi-etal-2007} 
{Daddi, E., et al.} 2007, \textit{ApJ}, 670, 156

\bibitem[Elsner et al.\ (2008)]{Elsner-etal-2008} 
{Elsner, F., Feulner, G, \& Hopp, U.} 2008, \textit{A\&A}, 477, 503

\bibitem[Fontana et al.\ (2006)]{Fontana-etal-2006} 
{Fontana, A., et al.} 2006, \textit{A\&A}, 459, 745

\bibitem[Gabasch et al.\ (2004)]{Gabasch-etal-2004} 
{Gabasch, A. et al.} 2004, \textit{A\&A}, 421, 41

\bibitem[Garn \& Best (2010)]{Garn-Best-2010}
{Garn, T. \& Best, P.N.} 2010, \textit{MNRAS}, 409, 421

\bibitem[Lee et al.\ (2010)]{Lee-etal-2010}{Lee, S.-K., Ferguson, H.C., Somerville, R.S., Wiklind, T., \& Giavalisco, M.} 2010, \textit{ApJ}, 725, 1644

\bibitem[Maraston et al.\ (2010)]{Maraston-etal-2010} {Maraston, C., et al.} 2010, \textit{MNRAS}, 407, 830

\bibitem[Perez-Gonzalez et al.\ (2008)]{PerezGonzalez-2008} 
{P\'erez-Gonz\'alez, P.G., et al.} 2008, \textit{ApJ}, 675, 234

\bibitem[Reddy \& Steidel (2009)]{Reddy-Steidel-2009}
{Reddy, N. \& Steidel, C.C.} 2009, \textit{ApJ}, 692, 778

\bibitem[Sawicki (2011a)]{Sawicki-2011a}
{Sawicki, M.} 2011a, \textit{MNRAS}, submitted

\bibitem[Sawicki (2011b)]{Sawicki-2011b}
{Sawicki, M.} 2011b, \textit{PASP}, in preparation

\bibitem[Sawicki et al.\ (2007)]{Sawicki-etal-2007} {Sawicki, M., et al.}\ 2007, in Astronomical Society of the Pacific Conference Series, ``Deepest Astronomical Surveys'', eds.\ J.\ Afonso, H.C.\ Ferguson, B.\ Mobasher, \& R.\ Norris, 433 

\bibitem[Sawicki \& Thompson (2006)]{Sawicki-Thompson-2006}
{Sawicki, M. \& Thompson, D.} 2006, \textit{ApJ}, 642, 653

%
\bibitem[Sawicki \& Yee (1998)]{Sawicki-Yee-1998}
{Sawicki, M. \& Yee, H.K.C.} 1998, \textit{AJ}, 115, 1329

\bibitem[Shapley et al.\ (2005)]{Shapley-etal-2005}
{Shapley, A.E., Steidel, C.C., Adelberger, K.L., \& Pettini, M.} 2001, \textit{ApJ}, 562, 95

\bibitem[Steidel et al.\ (1996)]{Steidel-etal-1996} {Steidel, C.C., Giavalisco, M., Dickinson, M., \& Adelberger, K.L.} 1996, \textit{AJ}, 112, 352 

\bibitem[Yabe et al.\ (2009)]{Yabe-etal-2009}
{Yabe, K., et al.} 2009, \textit{ApJ}, 593, 507



\end{thebibliography}
\end{document}